\documentclass{iau}
\usepackage{graphicx}
\usepackage{color}

\newcommand{\evi}{\mathcal E}

\newcommand{\brat}{\mathcal B}

\title[Searching for bias and correlations in a Bayesian way] 
{Searching for bias and correlations in a Bayesian way - Example: SN Ia data}

\author[Caroline Heneka, Alexandre Posada, Valerio Marra \& Luca Amendola]   
{\underline{Caroline Heneka}$^1$,
Alexandre Posada$^2$,  Valerio Marra$^{3}$ \and Luca Amendola$^4$}

\affiliation{$^1$Dark Cosmology Centre, Niels Bohr Institute, University of Copenhagen \\ Juliane Maries Vej 30,
DK-2100 Copenhagen, Denmark \\ email: {\tt caroline@dark-cosmology.dk} \\[\affilskip]
$^2$Centre de Physique th{\'e}orique, Universit{\'e} d'Aix-Marseille
 \\ Campus de Luminy Case 907, 13288 Marseille cedex 9, France \\[\affilskip]
$^3$ Instituto de F{\'i}sica, Universidade Federal do Rio de Janeiro \\ CEP 21941-972, Rio de Janeiro, RJ, Brazil
\\[\affilskip]
$^4$Institut f{\"u}r Theoretische Physik, Universit{\"a}t Heidelberg \\ Philosophenweg 16, D-69120 Heidelberg, Germany}

\pubyear{2014}
\volume{306}  
\pagerange{119--126}
\jname{Statistical Challenges in 21st Century Cosmology}
\editors{A. Heavens, A. Krone-Martins \& J.-L. Starck, eds.}
\begin{document}

\maketitle

\begin{abstract}
A range of Bayesian tools has become widely used in cosmological data treatment and parameter inference (see \cite[Kunz, Bassett \& Hlozek (2007)]{Kunz_etal07}, \cite[Trotta (2008)]{Trotta08}, \cite[Amendola, Marra \& Quartin (2013)]{Amendola_etal13}).  With increasingly big datasets and higher precision, tools that enable us to further enhance the accuracy of our measurements gain importance. Here we present an approach based on internal robustness, introduced in \cite[Amendola, Marra \& Quartin (2013)]{Amendola_etal13} and adopted in \cite[Heneka, Marra \& Amendola (2014)]{Heneka_etal14}, to identify biased subsets of data and hidden correlation in a model independent way.
\end{abstract}

\firstsection 
\section{Introduction and method}

\begin{figure}
  \includegraphics[width=0.35\textwidth]{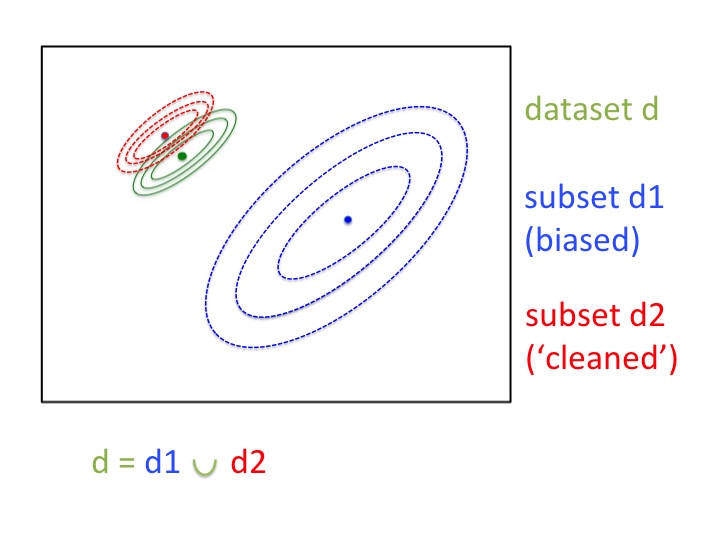} 
  \qquad
   \includegraphics[width=0.53\textwidth]{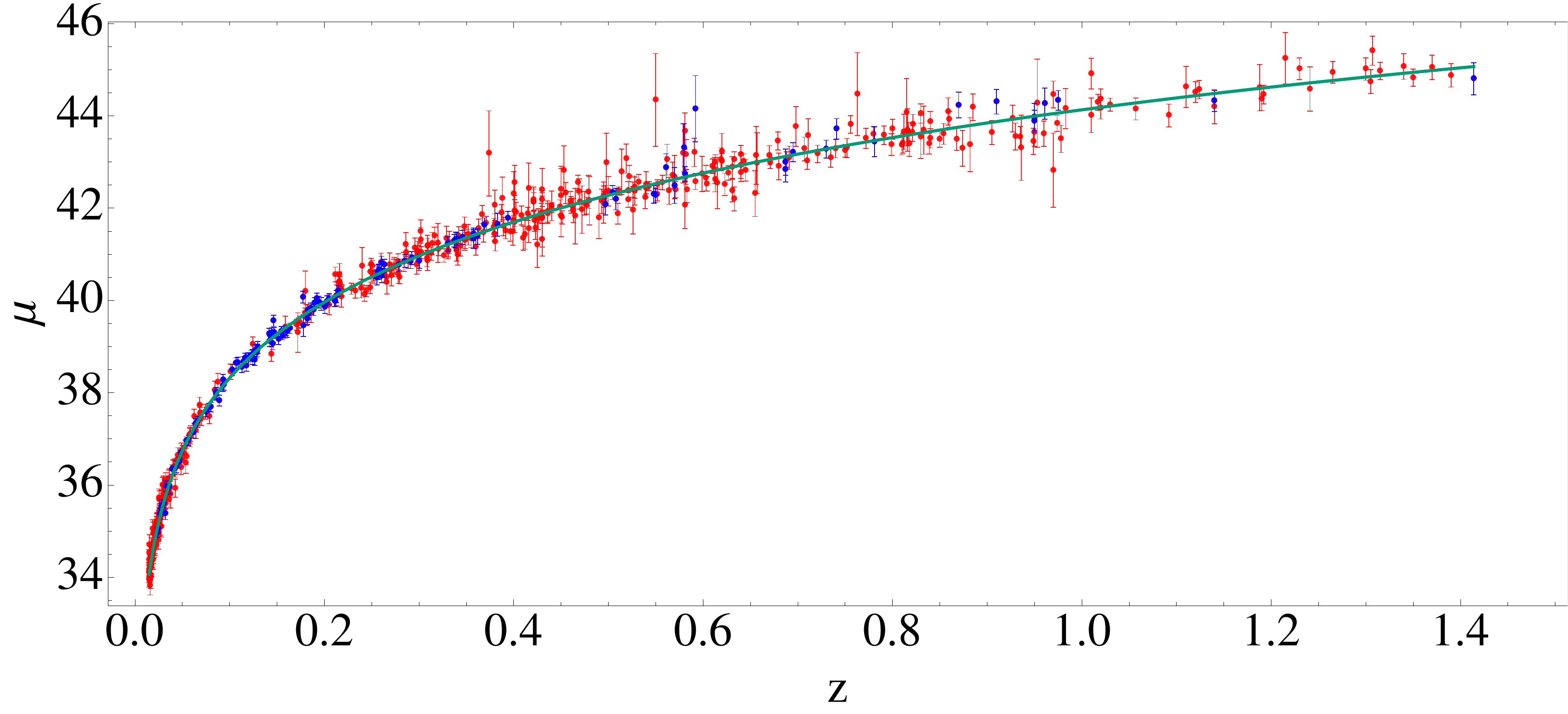} 
 \caption{Left: Sketch showing shift and change of size for likelihood contours when removing a biased subset ($d_1$) from the overall set (d). Right: Hubble diagram for the 580 SN Ia of the Union2.1 compilation (\cite[Suzuki et al. (2012)]{Suzuki12}),  best-fit cosmology in green, distance moduli with errors of subset of minimised robustness (R$\approx -283$) in blue, complementary set in red. Note that the otherwise indistinguishable biased set $d_1$ is identified.}
   \label{fig1}
\end{figure}

Our objective is the identification of subsets that differ from the overall data set in having a deviating underlying model. This deviation  becomes evident in form of a shift and change in size of likelihood contours (see 'biased' subset $d_1$ in blue as opposed to overall set $d$ in green in Fig.\ref{fig1}, left). Our method is useful for identifying deviating populations otherwise not distinguishable 'by eye' (see blue data points of lowest robustness in Fig.\ref{fig1}, right). We apply the formalism on supernova Ia data, the Union2.1 compilation (\cite[Suzuki et al. (2012)]{Suzuki12}) of 580 supernovae from z=0.015 to z=1.414. Observables are apparent magnitudes, stretch and colour corrected, as well as apparent magnitude errors.



\vspace{0.2cm}
{\underline{\it Internal Robustness Formalism}}.
We employ the Bayes' ratio to assess the compatibility between subsets statistically, making use of the full likelihood information. The hypothesis of having one model set of parameters $M_C$ to describe the overall dataset $d$ is compared with the hypothesis of having two independent distributions, i.e. parameter sets $M_C$ and $M_S$ for subsets $d_{1}$ and $d_{2}$. The corresponding Bayes' ratio of the evidences states, where we dub the logarithm of this ratio {\it internal robustness R}, 
\begin{equation}
\brat_{\rm tot,ind}=\frac{\evi\left(d;M_{C}\right)}{\evi\left(d_{1};M_{S}\right)\evi\left(d_{2};M_{C}\right)}  \hspace{0.2cm} and \hspace{0.2cm} R\equiv\log \brat_{\rm tot,ind}
\end{equation}
{\underline{\it Internal Robustness probability distribution function (iR-PDF)}}.
As an analytical form for the distribution of robustness values is not available, unbiased synthetic catalogues are necessary to test for the significance of the robustness values of the real data. They were created by randomising the best-fit function of the observable.
In practice we start by partitioning the data into subsets and choosing a parametrisation, followed by the evaluation of the robustness value for each partition. Finally, robustness values for real and synthetic catalogues can be compared to detect deviations, see Fig.\ref{fig2}.
\newline {\underline{\it Genetic Algorithm (GA)}}. We employ a genetic algorithm in order to find subsets of minimal robustness. Again, the parametrisation and initial subsets are chosen and their robustness values evaluated, followed by an iteration cycle of selection (in favour of subsets of lower robustness), reproduction (replacement of disfavoured with favoured subsets) and mutation (random data points are replaced) till convergence.

\section{Results}

\begin{figure}[h] 
\begin{center}
 \includegraphics[width=\textwidth]{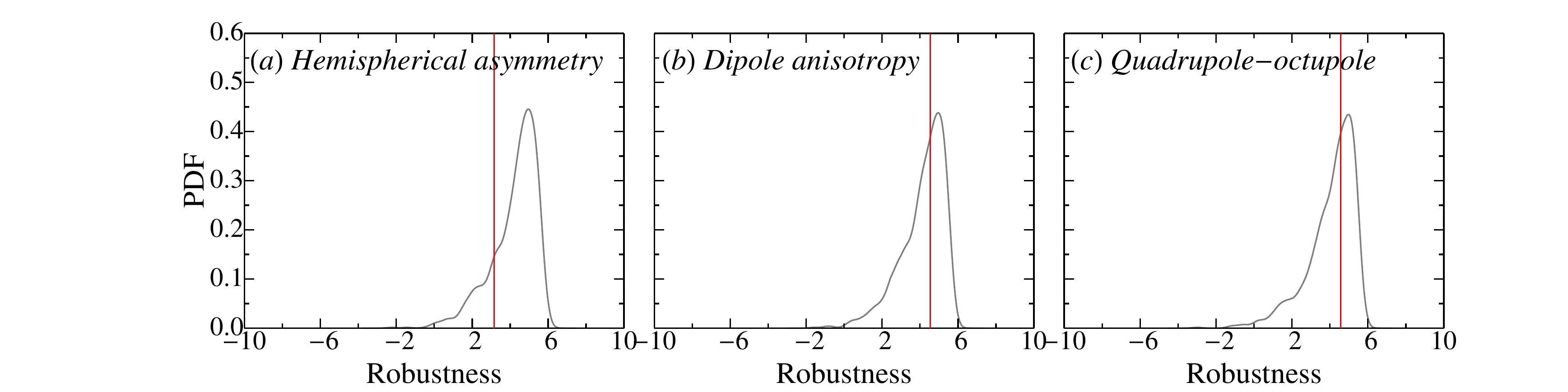} 
 \caption{Robustness test for three anisotropies reported by Planck: hemispherical asymmetry (left), dipole anisotropy (centre) and quadrupole-octupole alignment (right). The red vertical lines are robustness values of the Union2.1 Compilation, the distribution of the 1000 unbiased synthetic catalogues is shown in grey; taken from \cite[Heneka, Marra \& Amendola (2014)]{Heneka_etal14}.}
   \label{fig2}
\end{center}
\end{figure}

\begin{figure}[h] 
  \includegraphics[width=0.42\textwidth]{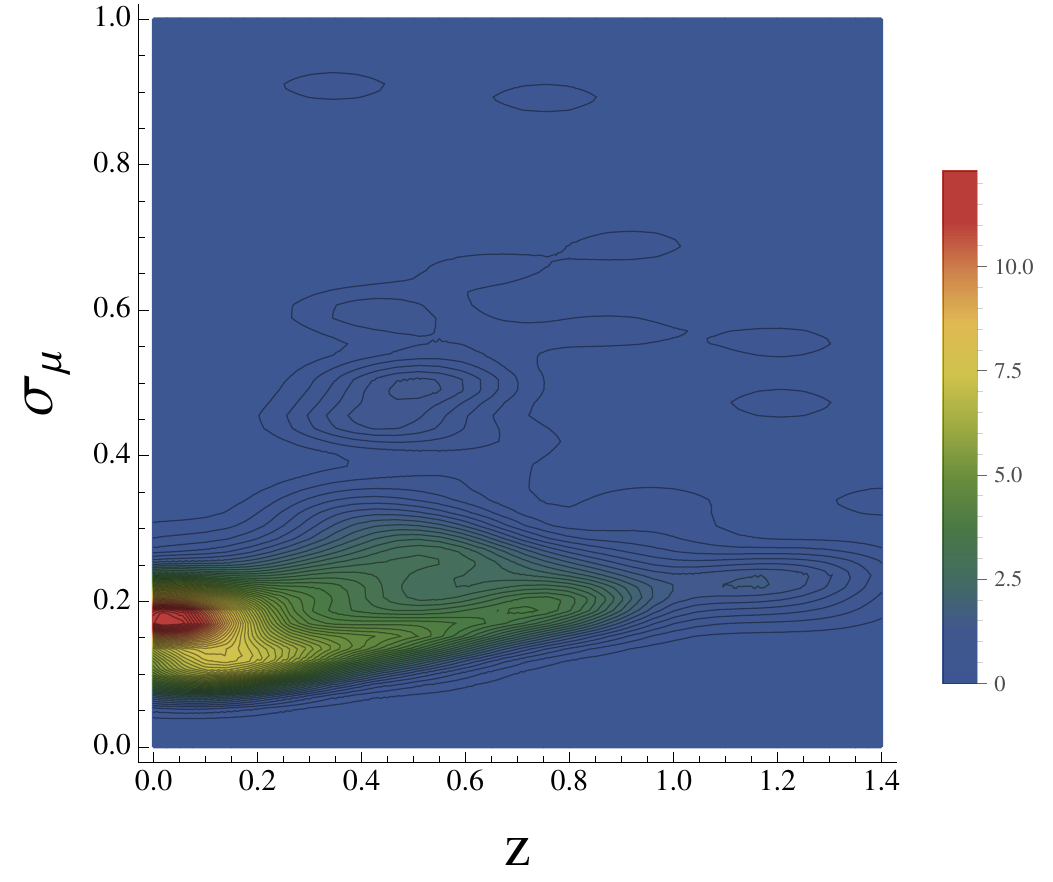} 
  \qquad
   \includegraphics[width=0.405\textwidth]{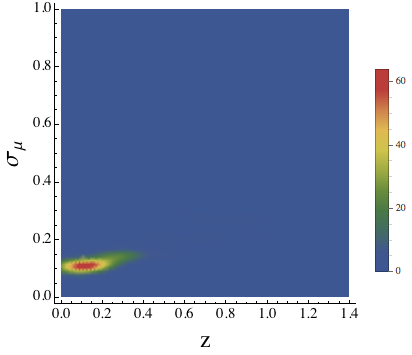} 
 \caption{Colour-coded contour plots for the occurrence of SN Ia in distance modulus error-redshift-space. Left:  Contour plot for a subset of $R\approx$-31, the subset of lowest robustness found for random $10^5$ subsets. Right: Contour-plot for the SN subset of minimal $R\approx$-283 found via GA. }
   \label{fig3}
\end{figure}

We employ the internal robustness formalism to search for statistically significant bias or correlations present in SN Ia data. The applicability to detect biased subsets, i.e. to identify subsets of deviating underlying best-fit parametrisation, is demonstrated. There are two ways to treat the data to form the iR-PDF: by randomly partitioning it into subsets to test in an unprejudiced way or by sorting the data after specific criteria to test prejudice on the occurrence of bias (for example angular separation, redshift, survey or hemisphere). Observables are both supernova Ia distance moduli and distance modulus errors. The tests of subsets partitioned due to certain prejudices showed no statistically significant deviation between real data and unbiased synthetic catalogues. This result demonstrates a successful removal of systematics for these cases and possible non-standard signals of anisotropy or inhomogeneity at only low level of significance. Fig.\ref{fig2} compares the iR-PDF of unbiased synthetic catalogues in grey with the real catalogue robustness value in red for anisotropies as reported by Planck (\cite[Ade et al., Planck Collaboration (2013)]{Ade_etal13}). For random partitioning subsets of low robustness can be identified. We show in Fig.\ref{fig3}, left panel,  the occurrence of distance modulus errors for the least robust set found by random selection.

The genetic algorithm (GA) randomly selects subsets for robustness analyses and transforms them due to selection rules in order to find subsets of minimal robustness. Seeking for the detection of systematics, distance modulus errors are analysed. 
Subsets of minimal robustness are found at  low values of $R\leq -280$.
Fig.\ref{fig3}, right panel, shows the occurrence of distance modulus errors with redshift for a subset of lowest robustness found via genetic algorithm minimisation.  Remarkably, most SNe found in these subsets occupy a confined region in distance modulus error - redshift - space and belong to distinct surveys of the overall compilation.

\section{Conclusion}
The applicability of the internal robustness formalism to detect subsets of data whose underlying model deviates significantly from the overall best-fit model is demonstrated. Subsets of lowest robustness for further investigation are identified, having higher probabilities of being biased. Both the degree to which systematics or cosmological signals unaccounted for are present can be quantified in an unprejudiced and model-independent way.  
This is crucial in order to detect contaminants or signals in cosmological or any astronomical data, especially with upcoming surveys  rendering a hunt for bias Ôby-handÕ more and more problematic.

\vspace{0.1cm}
\textbf{Acknowledgements.}
The Dark Cosmology Centre is funded by the DNRF. V.M. is supported by a Science Without Borders fellowship from the Brazilian Foundation for the Coordination of Improvement of Higher Education Personnel (CAPES). L.A. acknowledges support from DFG through the project TRR33 "The Dark Universe".


\begin{thebibliography}{}


\bibitem[Ade et al., Planck Collaboration (2013)]{Ade_etal13}
{Ade, P., et al. (Planck Collaboration)} 2013,
1303.5083

\bibitem[Amendola, Marra \& Quartin (2013)]{Amendola_etal13}
{Amendola, L., Marra, V.  \& Quartin, M.} 2013, 
\textit{MNRAS}, 430, 1867

\bibitem[Heneka, Marra \& Amendola (2014)]{Heneka_etal14}
{Heneka, C., Marra, V. \& Amendola, L.} 2014, 
\textit{MNRAS}, 439, 1855

\bibitem[Kunz, Bassett \& Hlozek (2007)]{Kunz_etal07}
{Kunz, M., Bassett, B.A. \& Hlozek, R.} 2007, 
\textit{Phys. Rev. D}, 75, 103508

\bibitem[Suzuki et al. (2012)]{Suzuki12}
{Suzuki, N.} 2012,
\textit{ApJ}, 746, 85

\bibitem[Trotta (2008)]{Trotta08}
{Trotta, R.} 2008,
\textit{Contemp. Phys.}, 49, 71

\end{thebibliography}
\end{document}